\documentclass[a4paper]{article}
\usepackage{graphicx}
\usepackage{amsmath}
\begin{document}

\begin{center}
{\bf \Large
Cooperation and defection in ghetto
}\\[5mm]

{\large
Krzysztof Ku{\l}akowski
}\\[3mm]

{\em
Faculty of Physics and Applied Computer Science,
AGH University of Science and Technology,
al. Mickiewicza 30, 30059 Krak\'ow, Poland
}

\bigskip
{\tt  kulakowski@novell.ftj.agh.edu.pl}

\bigskip
\today
\end{center}

\begin{abstract}
We consider ghetto as a community of people ruled against their will by an external power.
Members of the community feel that their laws are broken. However, attempts to leave ghetto
makes their situation worse. We discuss the relation of the ghetto inhabitants to the ruling 
power in context of their needs, organized according to the Maslow hierarchy. Decisions how to
satisfy successive needs are undertaken in cooperation with or defection the ruling power. 
This issue allows to construct the tree of decisions and to adopt the pruning technique from 
the game theory. Dynamics of decisions can be described within the formalism of fundamental 
equations. The result is that the strategy of defection is stabilized by the estimated payoff.
 
\end{abstract}



\section{Introduction}
In human history, violence is continuously with us, despite our optimistic belief 
that it is less and less widespread. In our minds, violence of armed against disarmed 
people is particularly repulsive. However, still it happens in numerous places on earth. 
Perhaps a new element is that all of us are more conscious of the situation, than ever 
before. The question is if victims of the violence - treated as a community - can accept 
it. If yes, the situation will remain stable; if not, they will resist, and the violence 
is expected to spread. This duality - to resist or not - is especially inevitable in a ghetto,
where an escape is not possible or at least very difficult. Here we are going to attack 
this problem by sociophysical methods, i.e. by a construction of an appropriate model.\\

As a model base, we propose two elements. First is the Maslow theory \cite{maslow}. 
Its basic assumption is 
that people are going to satisfy their needs one after another, in order from most 
basic to most sophisticated. Our central question - the decision of the victims about 
resistance - is to be considered in the context of their subsequent decisions how to 
satisfy their needs. In other words, each vital decision in the situation of violence 
is to be made in relation to this violence. Second element of the model is the mean field
approach, as applied to a strike by Galam, Gefen and Shapir \cite{galam}. This 
approach profits from the analogy to ferro-paramagnetic transition in the Ising model. In this
description, the ferromagnetic phase with given orientation of spins is equivalent to
a given decision (as to take part in a strike or not), made by the majority as a consequence
of social interactions. \\

This author imagines that the goal of this paper is twofold. First is a reconstruction 
of chains of subsequent decisions of people. We apply the decision tree - a concept from 
the game theory \cite{straffin}. This concept is modified here in the sense that 
there is only one player. Still, each decision selects a branch, and getting there 
defines a new situation. In some cases, estimation of expected payoffs allows to 
apply the pruning technique: as the player decides as to get larger payoff, some branches
with very low payoff can be {\it a priori} eliminated. The 
second goal is to use the obtained scheme to discuss the problem of resistance in a ghetto.  \\

Historically, "ghetto" is the area of iron foundry in Venice, where an enclosed 
neighborhood was created for Jews in 1516 to protect them against persecution from 
Roman Catholic Church. More recently, the term 'ghetto' is explicitly assigned to 
bounded areas in Warsaw, Lodz, Riga, Budapest, Cluj, Terezin and many others under 
Nazi rules, where Jewish people were gathered prior to the Holocaust. In social 
sciences, the meaning of the term includes also Jewish diaspora in early modern 
Europe, quarters of black Americans in large cities and some ethnic communities 
in Africa and East Asia \cite{smelser}. Although this meaning remains under 
dispute \cite{smelser}, attempts to describe ghettos in current world should at 
least be remarked \cite{hannerz, ron}, not pretending to completeness. Social
processes leading to the formation of ghettos was simulated by \cite{lling} and 
more recently by \cite{meyort,sch}. The present work concerns with dynamics of
decisions in a ghetto community. For our 
purposes, two traits are to be distinguished: {\it i)} an attempt of an inhabitant of 
ghetto to leave the area makes his situation worse, {\it ii)} human laws, as understood 
by inhabitants, are broken by an external power. This wide definition applies to 
refugee camps as well as settlements in countries controlled by army, as in 
Palestine, Tibet, Chechnya and Darfur. Although in most cases ghettos are inhabited 
by ethnic minorities, here we do not need to emphasize this ethnic trait.\\

To refer to the game theory, below we adopt the abbreviations $C$ and $D$ (cooperate or 
defect), although maybe withdraw or resist would be more appropriate. In two subsequent 
sections we introduce the model and we apply it to the case of ghetto. Last section is 
devoted to discussion.

\section{The model}
\subsection{Hierarchy of human needs}
According to the theory of Maslow, human needs are arranged hierarchically, from physiological 
needs, safety, belongingness, to esteem and self-actualization \cite{maslow}. People 
become interested in their safety to the extent of which their physiological 
needs are satisfied; being safe, they start to struggle for belongingness, and so on. 
This author and maybe this reader happened to be born in a milieu 
where three first needs were satisfied from the very beginning till adulthood. However, 
in numerous cases the situation is less fortuitous. More than often, a human unit has 
to determine a strategy to realize his/her most basic needs in this or that way. In 
such a strategy, one of most important decision is which limitations of human needs 
are to be accepted \cite{kepinski}. This problem appears to be even more crucial in 
ghetto, where the above mentioned limitations are particularly painful.\\

Trying to reach its needs in any milieu, a human unit has to consider at each step 
the context of situation. In particular, in ghetto the problem of any action is if 
it is legal, or - in other words - if this action is allowed by the ruling power. This 
remains true when we ask about actions taken up in order to satisfy human needs 
at all levels. At the physiological level, to cooperate is  equivalent to join common 
life in frames of the society, using money, sleeping home and eating food bought in 
a market. An alternative is to look for a desolate place in a forest or a desert, or 
to form a small community out of or at least at the border of, say, normal civilization. 
To continue, at the safety level the problem is to accept law or not. Whereas in our 
world of white collars this alternative is concentrated around payment of taxes, in 
ghetto the defection can include uprising, riots, guerilla or terror. At the level of 
belongingness, 
we have to select our group of reference. Again, in ghetto the world is sharply divided 
into two: "we" and "they". The power is with "them", and the quest is to identify with 
whom? Further, at the level of esteem the problem is, in which group this esteem is 
looked for? Here we guess that this choice is strongly correlated with the previous 
one, and our analysis will be simpler because of this correlation. Finally, reaching 
esteem, our human unit tends to self-actualization. In principle, this again can be 
expressed as a social action directed against the power or supporting it. However, 
consequences of these decisions are usually less crucial, human behavior at this 
level is much more individualized and it is often affective and expressive rather 
than aim-oriented \cite{weber}. That is why the level of self-actualization is 
difficult for a sociophysical modelling.

\subsection{Tree of decisions}
Those who defect at the very root, i.e. at the physiological level, place themselves 
out of frames of the society. It is very hard to defect public access to water, shops 
and houses. The alternative is to live wildly in the forest. Yet this choice happens 
in several places on Earth, where climatic circumstances allow to do it at least 
temporarily and some strong obstacles prevent to live in accord with law.  This 
kind of defection happens in large social scale only in societies in strongest crisis, 
e.g. during a civil war. Although these situations are of central importance, they 
will be not discussed here. In a ghetto, there is no possibility to fight in open 
way; main splitting of human behavior happens at higher levels. Then, for the sake 
of our subject all decisions discussed here start from $C$ (cooperate).\\

In the same way we are going to comment further decisions, which form chains as the 
one presented in Table 1. This particular chain will be denoted as $CCDDD$ from now on. 
In this notation, $CD$ means that we discuss the decision $D$ (at the safety level) of 
those who decided $C$ at the physiological level. \\
\bigskip

\begin{tabular}{|l|l|}
\hline
physiology&C\\
safety&C\\
belongingness&D\\
esteem&D\\
self-actualization&D\\
\hline
\end{tabular}

\bigskip

Table 1. A chain of subsequent decisions of a human unit (man or woman): 
defect ($D$) or cooperate ($C$) with the power.  

\bigskip
In Fig.1 a part of the resulting tree of decisions is shown. There, the whole branch 
starting from $D$ at the physiological level is omitted, except its beginning. Also,
the decisions $D$ or $C$ at the level of self-actualization are not shown for clarity
of the figure. \\

Omitting the physiological needs, we are going to concentrate on the level of safety.
A population considered selected $C$ as their first choice, i.e. they decided to live
within the community and to profit its facilities. Now and each time their decision 
is $C$ or $D$, i.e. they wonder if their path is to be $CC$ or $CD$. The probabilities of these
paths depend on the expected payoffs. Then the choice these people is to decide, if they 
will be safer when cooperating with the power of when defecting it.\\

\begin{figure} 
\vspace{0.3cm} 
{\par\centering \resizebox*{10cm}{8cm}{\rotatebox{0}{\includegraphics{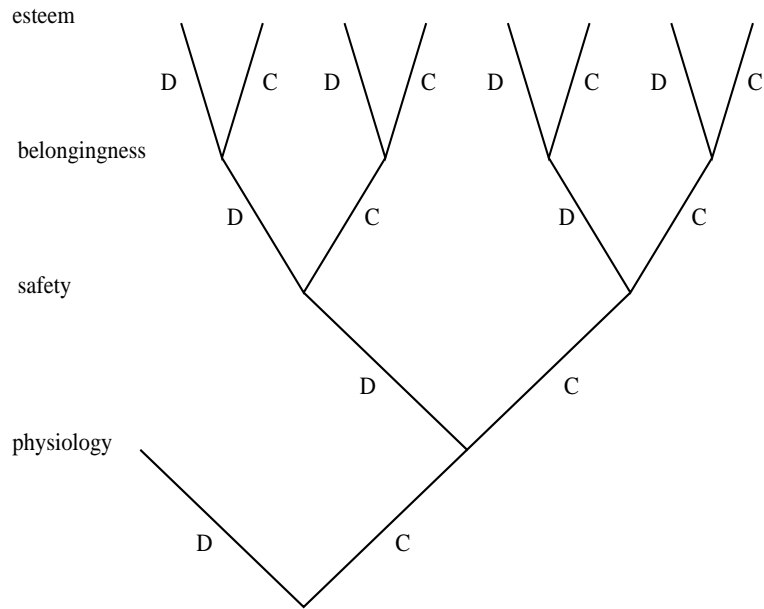}}} \par} 
\vspace{0.3cm} 
\caption{Right half of the tree of decisions. Last level (self-actualization) is not shown.
Being at the root and selecting $C$, one is placed at node $C$; selecting $D$ as next level one 
is placed at node $CD$ and so on. Then, nodes at the physiology level are indexed with one 
label, $D$ and $C$ from left to right, nodes at the safety level - by two labels ($DD$ or 
$CD$ not shown, or $CD$ or $CC$ shown from left to right), and so on. At the esteem level,
first node from the left is indexed as $CDDD$.}  
\end{figure}

As it was indicated by Maslow, people are able to struggle for their safety
to the extent in which their physiological needs are satisfied. Further, their search for 
belongingness is limited by their lack of safety, and so on. Maslow gives an example 
with numbers; an average citizen could have satisfied his successive needs in 85, 70,
50, 40 and 10 percent, in order as in Table 1. Provided, for example, that the safety needs 
of somebody are not satisfied at all, he will not bother about belongingness, not to speak about 
esteem or self-actualization. It is not clear how the effort for a need of next level 
depends on the satisfaction of a need in a previous level; the Maslow theory is 
formulated in words, not in numbers. The area is open for speculations, with the only condition 
that any proposed mathematical formulation reflects the above mentioned rules. On the other hand 
it is obvious that the validity
of any numbers we can get is limited to statistical considerations. It seems that
for this kind of problems, the fundamental equations \cite{vank} can provide a proper tool. \\

\subsection{Mathematical formulation}

From these equations,
we expect to obtain the probabilities that people in a given situation (read: at a given
node of the tree) select this or that decision. External conditions met by a given community can be
introduced as the set of payoffs $\alpha_X$, describing maximal possible percentage of satisfying
needs at node $X$ of the decision tree. As explained in the caption to Fig. 1, the node index $X$ 
is equivalent to a chain of decisions, leading to that node.
The root is treated as the chain of zeroth length. Simultaneously, $\alpha_X$ is 
the maximal amount of people who are able to struggle for satisfaction of higher needs at nodes
$XC$ and $XD$. Both these 'maximal' deal with a virtual case when the payoff is limited neither by 
parameters of previous nodes, nor by human decisions at these nodes.\\

Keeping the above example of the chain $CCDDD$ as an individual path, 
the value of satisfaction $s_X$ of a human unit - member of the 
community - at node $C$ (at the physiological level) is then $s_C=\alpha_C$. At higher level $s_X$
fulfils an iterative equation

\begin{equation}
s_{Y}=s_X \alpha_{Y},
\end{equation}
for $Y=XC,XD$. Provided, that the set $s_X$ of satisfaction of our human unit accords with 
the above exemplary values given by Maslow, we obtain at five successive nodes of the path 
$\alpha_C=0.85$, $\alpha_{CC}=0.7/0.85\approx 0.82$, $\alpha_{CCD}=0.5/0.7\approx 0.71$, 
$\alpha_{CCDD}=0.4/0.5=0.8$ and $\alpha_{CCDDD}=0.1/0.4=0.25$. These values of $\alpha_X$
allow to reproduce {\it via} Eq. 1 the above given exemplary chain of individual satisfactions: 
$s_C=0.85$, $s_{CC}=0.7$, $s_{CCD}=0.5$, $s_{CCDD}=0.4$ and $s_{CCDDD}=0.1$. \\

The above vital path consists successive decisions, for our example $CCDDD$, 
as in Table 1. In reality, these decisions are much more detailed than a cooperation
with or a defection the ruling power. Actually, the decision can be as specific as to marry
one particular member of a group of revolutionists - or not to marry. However, having defined our 
issue - to withdraw or to resist, we are interested not as in a decision of selecting a detailed person, but - averaging out
over different possibilities - in a decision to be involved in a revolutionistic group, which
is equivalent to satisfying some needs by the choice $D$. \\

Up to now, we dealt with individual path. Now we can introduce conditional probabilities
$p(C\mid X)=p(XC)/p(X)$ and $p(D\mid X)=p(XD)/p(X)$ that, leaving node $X$, a human unit is going 
to $C$ or $D$.
In this case the normalization condition should be $p(C\mid X)+p(D\mid X)+1-\alpha_X=1$. This 
should be not 
confused with a probability that a human unit will stay at node X with probability $1-\alpha_X$.
Such a formulation would disagree with the original interpretation of Maslow. Instead, the factor
$1-\alpha_X$ measures the amount of effort spent inefficiently at node $X$, in the same way as 
it was
assumed for an individual path. In the latter case, either $p(C\mid X)=0$ or $p(D\mid X)=0$, 
and the path
was fully determined by subsequent individual decisions. Then, individual satisfaction at 
subsequent levels depends only on the payoffs $\alpha_X$, as explained in the second paragraph 
of this section. Instead of using the conditional probabilities, it is simpler to use individual 
effort $w_X$ and averaged effort $W_X$. In the above example of individual path, $CCDDD$, the 
set of individual efforts is: $w_C=1$ at the root, $w_{CC}=1$ at the node $C$, $w_{CCD}=1$ at
the node $CC$, $w_{CCDD}=1$ at the node $CCD$ and $w_{CCDDD}=1$ at the node $CCDD$. Other efforts
are zero, either along the decision (as $w_{CD}$) or because a given node was not reached
by a given human unit (as $w_{DD})$. \\

Averaging over individual paths, we get a set of average amounts of effort $W_X$ 
at all nodes $X$. Then for the physiological level we have $W_C+W_D=1$. The average satisfactions
at the physiological level, $X=C,D$, are $S_C=W_C\alpha_C$, and $S_D=W_D\alpha_D$.  Considering 
the safety level we take into account that efforts to reach the nodes $CC$ and $CD$ 
are reduced because $\alpha_C\le 1$. Then, $W_{CC}+W_{CD}=W_C\alpha_C$, and similarly 
$W_{DC}+W_{DD}=W_D\alpha_D$. As a rule,
  
\begin{equation}
W_{XC}+W_{XD}=W_X\alpha_X,
\end{equation}
where $XC$ and $XD$ are nodes available from node $X$ by decision $C$ or $D$.
The whole set $W_X$ is equivalent to a map of social efforts, put into various ways of attempts
of satisfying the needs. At each node, the average satisfaction $S_X=W_X\alpha_X\le W_X$.
Satisfaction is less or equal than effort, for individual paths as well as in the average.\\

In a deterministic picture, people are expected to select always the nodes with larger payoff. 
However, it is clear even for a physicist that in reality people have their individual preferences,
and a common payoff for everybody can be introduced only for a statistical description. This
intuition on individual character of payoffs is confirmed by the utility theory \cite{straffin}.
Working in statistical physics, we are tempted to use some noise as a measure of, say, lack
of information of the community members. Then we expect that the ratio $W_{XC}/W_{XD}$ in stationary 
state depends on $\beta(\alpha_{XC}-\alpha_{XD})$, where $\beta=0$ for absolute lack of information
on the payoffs, and $\beta$ is large when the information is well accessible. From this point, it 
is only one step to mimic the statistical mechanics, writing the stationary probability of 
selecting $C$ from node $X$

\begin{equation}
p(XC)_{eq}=\frac{W_{XC}}{\alpha_X W_X}\propto \exp{[\beta(\alpha_{XC}-\alpha_{XD})]},
\end{equation}
and to postulate a dynamic description in the form of fundamental (or Master) equation

\begin{equation}
\frac{dp(XC,t)}{dt}=-r(XC)p(XC,t)+r(XD)[1-p(XC,t)],
\end{equation}
where $r(XC)\propto p(XC)_{eq}$. Here, the constant of proportionality determines the 
timescale of the dynamics. The dynamics of the probabilities $p(X)$ is equivalent to the dynamics 
of efforts $W_X$.\\

In, say, a standard society the information on the payoffs is well accessible and the 
successive selections are almost deterministic. Then, people who decide to live in a wild forest 
are rare exceptions in the society: almost everybody selects $C$ at the physiological level.
It is less clear if the payoffs for those who break law are indeed smaller than for the others.
In any case, a great effort is paid to ensure the population that sooner or later this
payoff will be strongly reduced. Because of this effort, the statistical data on the choice 
of $CD$ are usually less sure. Looking for belongingness and needs of higher order is not directly 
connected with our issue; anyway, in democratic systems we are partially involved
into the ruling power, which cannot then be treated as external and is maybe not entirely against 
our will. Summarizing this section, this author believes that the concept of the decision tree,
as an adaptation of the Maslow hierarchy, can be useful in many issues.

\section{The case of ghetto}

As it is expected to be clear from the definition of ghetto, accepted above, the key point
of the decision tree is the node $C$, where crucial decision is to be taken: $CC$ or $CD$. The reason 
is as follows. All what we know about ghetto confirms that there, it is almost impossible to 
satisfy the safety need. The payoff of a useful solution $CC$ is drastically reduced with respect
to other communities. Examples of this painful truth fill newspapers and TV or, even worse,
remain unknown if information is prohibited. To bring these examples here, athough justified
from the point of view of the subject, would drive us too far from sociophysics. Instead, let us 
consider the consequences for the payoff.\\

Imagine that the safety is strongly reduced in an initially normal society. The reason can be 
war or revolution, or other abrupt fall of the political system.  It is clear that the accessibility
of information deteriorates, and in this situation many people do not know what to do. What
is the payoff if I withdraw? if I resist? who will win? what will be the consequences for me?
my family? my assets? and so on. As a rule, a remarkable percentage of people resist, just
because - a physicist would say - large entropy in the system. This thermodynamic 
formulation should not be offending to anyone. Obviously, it does not comprise individual 
decisions, which are sometimes dramatic and full of unanswered questions. It is a common 
experience that we
decide, not knowing the final results; in most difficult situations, the amount of information
is too low to allow for a logical reasoning. This experience is encoded in sociology as the law 
of unforeseen consequences \cite{oxf}. However, here we consider the case when finally 
some power, external for the ghetto inhabitants, prevails and the information on payoffs becomes 
more clear. But the above mentioned 
group keeps resisting; despite the variety of their motivations, their effort can be translated 
into numbers and handled by statistical tools.
 They fight against the external power and its supporters - the mechanism
known too well, indeed. Relaxing to the stationary state, the system finds that the payoff
of the choice $CC$ is reduced by an expected repression by the resisting group. The ruling
power tries to balance this repression by defeating the resistance fighters. Soon, the level of
aggression of both sides becomes equivalent; both find convenient justifications. \\

This author believes that what can be said mathematically, can be said - although longer - in words.
Here we try the opposite way. Violence bears violence - this sentence is short. In sociophysical
language, the same content can be expressed as a stability of the solution of Eq. 4, characterized
by the condition $\alpha_{CC}<\alpha_{CD}$. This stability relies on the following premises: 
{\it i)} the payoff $\alpha_{CC}$ is drastically lowered by the repressive actions of the resistant
group, {\it ii)} struggling for their safety, people are not motivated to select $CC$ instead of 
$CD$, if $\alpha_{CC}<\alpha_{CD}$, {\it iii)} selection of $CD$ in a social scale reinforces
the resistant group. As we see, this closed circle does not rely on a particular choice of the
functional dependence of the effort $w$ on the payoff $\alpha$. In fact, the resistant group
can be compared to a nucleation center, which initiates the new phase. However, the nucleation
process cannot be described within the simplest version of the mean field theory, used here.\\

As a result, the whole tree becomes degenerated. For those who decided to resist, it is not possible 
to look for belongingness or esteem out of the resistant group. On the other hand, those who 
select $CC$ remain under fear of, from one side, being accused of treason and, from the other side, 
blind actions of the ruling power. Not being able to get safety, they follow
the solidarity with the resistant group, whey they look for belongingness and esteem. As a rule, 
when the safety is at risk, no effort can be put to struggle for higher needs. In effect,
upper branches disappear. \\

Trying to illustrate the above processes with some simulations, we need the values of several 
parameters, as the payoffs at the nodes etc. Measurement of these parameters or at least a thorough
discussion of their values far excesses the frames of this work - in social sciences, this is 
almost an euphemism. Instead, we can present
a qualitative consequences of an abrupt change of ruling power. The event is a special and 
most simple example of what was discussed before. We limit the calculations to the safety level. 
In the formalism developed above, the dynamics of this level does not depend on the parameters 
on higher levels. In the calculations, the difference of the payoffs consists two factors:
external $\Delta$ (provided by the ruling power, old or new) and internal, due to interaction 
between the community members. The latter is proportional to the actual difference of efforts,
$W_{CD}-W_{CC}$. This proportionality encodes the above discussed positive feedback between
the value of the difference of efforts and its time derivative. In fact, this positive feedback 
is at the core of the mean field theory \cite{binder}.\\

To simulate the change of the ruling power (for example, from a well-established to an 
external), two agents cannot be omitted: a strong decrease of $\alpha_C$, which is a direct 
consequence of unavoidable war, and a change of sign of $\Delta$. Simultaneously, the up-to-now
cooperators become defectors and the opposite. We keep the node $D$ unoccupied ($W_C=1$ and $W_D=0$); 
this reflects
the assumed fact that nobody can leave the ghetto. For simplicity we keep the parameter $\beta$
constant in time, although it is almost surely not realistic; still we are left with three parameters.
The value of $\alpha_C$ before the political overthrow is assumed to be unity. Its value after 
the overthrow, kept constant in time, is one of the parameters. The remaining parameters are
$\beta$ and $\Delta$. Initial ratio of the variables $W_{CC}$ and $W_{CD}$ is taken as their ratio 
at equilibrium before the overthrow. As a rule, $W_{CD}>W_{CC}$ at initial time, because most 
people supported the ancient regime before the overthrow; what was the cooperation, now is treated 
as defection, and the opposite.

\begin{figure} 
\vspace{0.3cm} 
{\par\centering \resizebox*{10cm}{8cm}{\rotatebox{270}{\includegraphics{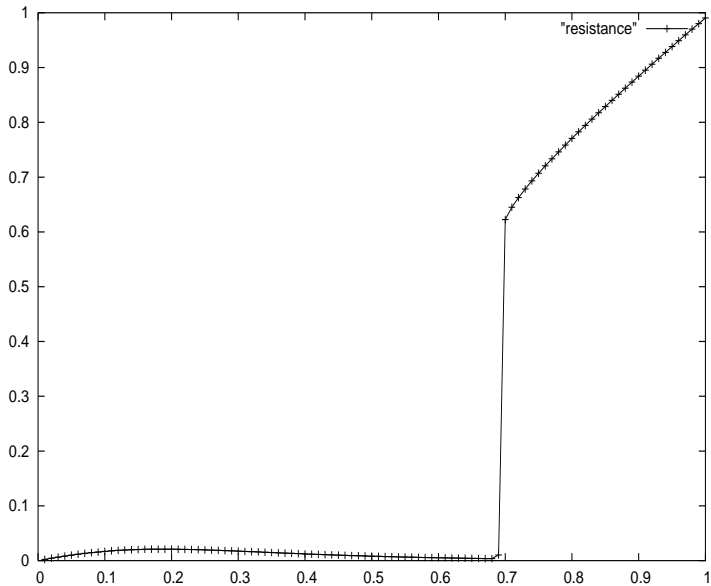}}} \par} 
\vspace{0.3cm} 
\caption{The effort $W_{CD}$ put at the resistance at the safety level against the parameter
$\alpha_C$.}  
\end{figure} 

In Fig. 1 we show the effort $W_{CD}$ put at the resistance, against the satisfaction $\alpha_C$
of the physiological needs at node C. These data are for the stationary state. As remarked above, 
we assume that all the social effort
at the root is put to satisfy physiological needs within the community. However, these needs can
be satisfied only partially. The parameter $\alpha_C$ measures the level of this satisfaction. 
Further, it measures also the effort which can be put to struggle for safety, in this ($CC$) or that
($CD$) way. The parameters for the plot are: $\beta=3.0$, and $\Delta=0.2$.  \\

As we see in Fig. 1, there is a jump of the data on $W_{CD}$ near $\alpha_C=0.7$. Below this value, 
the effort put to resistance is negligible. Above this value, it is close to its maximal value 
$\alpha_C$. This means that the initial state of resistance is stable. The results are typical,
as long as $\beta$ is not too small, and $\Delta$ is not too large. Within the magnetic analogy,
the results mean that the metastable phase is possible as long as the field ($\Delta$) and the 
temperature ($1/\beta$) are not too large. Within the sociophysical picture, it means that it is 
advantageous for the ruling power to keep the whole ghetto community at the limit of starvation
i.e. with small value of satisfaction $\alpha_C$ of physiological needs. Then, instead of fighting, 
they are kept in a queue for water and flour, provided by the army. Then, the best thing is to make 
a movie and show it in TV news; those who get water are happy. Please do not blame this author for
the invention - it is known for a long time.

\section{Discussion}

Our conclusions are to be divided in three parts. The first is sociophysical. Our mathematical 
description is equivalent to the mean-field theory of the ferromagnetic phase, where
two stable solutions coexist \cite{binder}. This model is well established in applications
of physics to social sciences \cite{galam,weidlich}. It is known that the stability of the 
ferromagnetic
phase is overestimated by mean field theory; in fact, it depends on the structure. Here
we are faced with the question, what is a realistic structure of a community. Much effort has been
done by sociologists to advance our knowledge on the subject; however, even the characteristic 
size of social networks remains under dispute \cite{kill,mars}. On the other hand, stability
of ordered Ising phase at low temperatures has been found in computer simulations for most of 
investigated structures \cite{holyst,tadic,makowiec,malarz}, with directed Albert-Barab\'asi 
networks \cite{sum,lima} and one-dimensional chains and related 
models \cite{novot} as exceptions. Actually, time dependence of persisting opinion
of a resistant group was discussed recently by \cite{aydiner} on the basis of one-dimensional Ising model. 
(We note that the condition of low temperature
is equivalent to large value of the uncertainty factor $\beta$ in our considerations.) For social 
applications, the condition of an eternal stability of the ferromagnetic phase can be 
substituted by a weaker condition of appearance of long-living ordered domains. We can conclude 
that sociophysical arguments work for this hypothesis, and not against it.\\

The second conclusion is aimed to be sociological. The results of our analysis indicate, that when
an external power struggles for control of an isolated community, 
the problem of safety remains crucial. Obvious aim of the power must be: to guarantee the safety
for still neutral part of the community. If this is not possible, war becomes eternal, without
winners. Not so rarely, the responsability for safety of isolated communities remains in hands 
of army, without
control of civil agencies or free press. This is precisely what eliminates the possibility
of a peaceful solution; army people are trained to fight, not to bring safety.
In sociology, the role of safety is known for a long time: first edition of the Maslow's
book \cite{maslow} appeared in 1954. The advantage of this work is to express it in more
formalized language. One could ask if such a formulation is worthwhile. On the other hand,
still some powerful people seem not to recognize the validity of the conclusions of Maslow theory.
Maybe they will be convinced by mathematics.\\

In my last word I declare to share the opinion that ghettos are shameful for human 
civilisation. Nevertheless this respectable and rather common opinion, such places exist, 
as we are mercilessly informed by free media. Some people even claim, that some 
of these places are established to protect our laws to free life, where at the last level 
of the tree of decisions we can write our sociophysical papers. If this is done without
care about safety of the ghetto inhabitants, the way is destructive and mindless. \\

{\bf Acknowledgements.} The author thanks Ma{\l}gorzata Krawczyk and Francis Tuerlinckx 
for their kind help. Thanks are due also to Dietrich Stauffer for helpful criticism and 
reference data.

\end{document}